\documentclass[10pt,twocolumn,amsmath,amssymb,floatfix,superscriptaddress]{revtex4-2}
\usepackage[utf8]{inputenc}
\usepackage{float}
\usepackage{graphicx}
\usepackage{dcolumn}
\usepackage{braket}
\usepackage{bm}
\usepackage{color}
\usepackage{txfonts}
\usepackage{microtype}
\usepackage{mathrsfs}
\usepackage{amsmath}
\usepackage{extarrows}
\usepackage{mathtools}
\usepackage{xcolor}
\usepackage{slashed}
\usepackage{comment}
\usepackage{paracol}

\usepackage[colorlinks=true, linkcolor=blue, citecolor=blue, CJKbookmarks=true,urlcolor=blue]{hyperref}
\begin{document}

\title{Generating Vector-Vortex $\gamma$ Photons by Nonlinear Compton Scattering}

\author{Yong-Zheng Ren}
\affiliation{Ministry of Education Key Laboratory for Nonequilibrium Synthesis and Modulation of Condensed Matter, State Key Laboratory of Electrical Insulation and Power Equipment, Shaanxi Province Key Laboratory of Quantum Information and Quantum Optoelectronic Devices, School of Physics, Xi’an Jiaotong University, Xi’an 710049, China}
\author{Mamutjan Ababekri}
\email{mamutjan@xjtu.edu.cn}
\affiliation{Ministry of Education Key Laboratory for Nonequilibrium Synthesis and Modulation of Condensed Matter, State Key Laboratory of Electrical Insulation and Power Equipment, Shaanxi Province Key Laboratory of Quantum Information and Quantum Optoelectronic Devices, School of Physics, Xi’an Jiaotong University, Xi’an 710049, China}
\author{Jun-Lin Zhou}
\affiliation{Ministry of Education Key Laboratory for Nonequilibrium Synthesis and Modulation of Condensed Matter, State Key Laboratory of Electrical Insulation and Power Equipment, Shaanxi Province Key Laboratory of Quantum Information and Quantum Optoelectronic Devices, School of Physics, Xi’an Jiaotong University, Xi’an 710049, China}
\author{Feng Wan}
\affiliation{Ministry of Education Key Laboratory for Nonequilibrium Synthesis and Modulation of Condensed Matter, State Key Laboratory of Electrical Insulation and Power Equipment, Shaanxi Province Key Laboratory of Quantum Information and Quantum Optoelectronic Devices, School of Physics, Xi’an Jiaotong University, Xi’an 710049, China}
\author{Qian Zhao}
\affiliation{Ministry of Education Key Laboratory for Nonequilibrium Synthesis and Modulation of Condensed Matter, State Key Laboratory of Electrical Insulation and Power Equipment, Shaanxi Province Key Laboratory of Quantum Information and Quantum Optoelectronic Devices, School of Physics, Xi’an Jiaotong University, Xi’an 710049, China}
\author{Zhong-Peng Li}
\affiliation{Ministry of Education Key Laboratory for Nonequilibrium Synthesis and Modulation of Condensed Matter, State Key Laboratory of Electrical Insulation and Power Equipment, Shaanxi Province Key Laboratory of Quantum Information and Quantum Optoelectronic Devices, School of Physics, Xi’an Jiaotong University, Xi’an 710049, China}
\author{Kun Xue}
\affiliation{Ministry of Education Key Laboratory for Nonequilibrium Synthesis and Modulation of Condensed Matter, State Key Laboratory of Electrical Insulation and Power Equipment, Shaanxi Province Key Laboratory of Quantum Information and Quantum Optoelectronic Devices, School of Physics, Xi’an Jiaotong University, Xi’an 710049, China}
\author{Ya-Qing Huang}
\affiliation{Ministry of Education Key Laboratory for Nonequilibrium Synthesis and Modulation of Condensed Matter, State Key Laboratory of Electrical Insulation and Power Equipment, Shaanxi Province Key Laboratory of Quantum Information and Quantum Optoelectronic Devices, School of Physics, Xi’an Jiaotong University, Xi’an 710049, China}
\author{Zhao-Hui Chen}
\affiliation{Ministry of Education Key Laboratory for Nonequilibrium Synthesis and Modulation of Condensed Matter, State Key Laboratory of Electrical Insulation and Power Equipment, Shaanxi Province Key Laboratory of Quantum Information and Quantum Optoelectronic Devices, School of Physics, Xi’an Jiaotong University, Xi’an 710049, China}
\author{Zhong-Feng Xu}
\affiliation{Ministry of Education Key Laboratory for Nonequilibrium Synthesis and Modulation of Condensed Matter, State Key Laboratory of Electrical Insulation and Power Equipment, Shaanxi Province Key Laboratory of Quantum Information and Quantum Optoelectronic Devices, School of Physics, Xi’an Jiaotong University, Xi’an 710049, China}
\author{Jian-Xing Li}
\email{jianxing@xjtu.edu.cn}
\affiliation{Ministry of Education Key Laboratory for Nonequilibrium Synthesis and Modulation of Condensed Matter, State Key Laboratory of Electrical Insulation and Power Equipment, Shaanxi Province Key Laboratory of Quantum Information and Quantum Optoelectronic Devices, School of Physics, Xi’an Jiaotong University, Xi’an 710049, China}
\affiliation{Department of Nuclear Physics, China Institute of Atomic Energy, P.O. Box 275(7), Beijing 102413, China}
\date{\today}

\begin{abstract}

Vector-vortex photons, characterized by a nonseparable coupling between polarization and orbital angular momentum (OAM), offer opportunities for optical manipulation, quantum communication, nuclear photonics, etc. However, their generation in the $\gamma$-ray regime remains challenging. Here, we put forward a novel method to generate vector-vortex $\gamma$ photons via nonlinear Compton scattering in elliptically polarized laser pulses. We reveal that tailoring laser ellipticity directs the multiphoton absorption to coherently populate OAM modes with opposite winding numbers, $\pm \ell$, tied to orthogonal circular polarizations, producing nonseparable spin--OAM photon states. For a linearly polarized laser of moderate intensity (dimensionless amplitude $a_0 \sim 1$), the mode-pair concurrence—a 0-to-1 measure of spin–OAM entanglement—reaches unity for MeV $\gamma$ photons, realizing maximally nonseparable radial- or azimuthal-type vector-vortex states. The laser amplitude further controls the accessible OAM spectrum.  
Our method offers a route to MeV vector-vortex photons, opening a new avenue for nuclear-scale structured photonics and high-energy quantum information.

\end{abstract}

\maketitle

Vector-vortex photons are structured light states characterized by a nonseparable coupling between polarization and orbital angular momentum (OAM), giving rise to spatially inhomogeneous polarization textures 
\cite{zhan2009:cylindrical,milione2011:higher,d2016:entangled,Forbes:2021tpp,forbes2025progress}. In the paraxial limit, a representative cylindrical mode of order $m$ is defined as $|V_{m,\delta}\rangle = (|R,m\rangle + e^{i\delta}|L,-m\rangle)/\sqrt{2}$, where $|R,m\rangle$ and $|L,-m\rangle$ denote right-circularly polarized (RCP) and left-circularly polarized (LCP) scalar vortex states, yielding radial-type ($\delta = 0$) or azimuthal-type ($\delta = \pi$) fields \cite{milione2011:higher,d2016:entangled}. 
At the single-photon level, this nonseparability manifests as intraparticle entanglement between helicity and OAM, whereas scalar vortex modes have separable polarization and spatial structure \cite{Allen:1992zz,Molina-Terriza:2007ydx,bliokh:2015spin,Knyazev_2018,forbes2024:orbital}. 
Structured light has enabled significant advances in optical manipulation \cite{shen:2019optical,Gc:2003prl,schulz2020:generalized}, quantum optics \cite{Mair2001Nat,Leach2009pote,Robert:2012sci,castellucci2021atomic}, and imaging \cite{Swartzlander2001optl,Swartzlander2008opte,segawa2014:resolution,liu2022:super}, with vector-vortex states providing a basis for high-dimensional spin--OAM state engineering \cite{dambrosio2012:complete,wang2016:advances,ndagano2017:creation,chen2020vector}. 
At \(\gamma\)-ray energies, such states could extend studies of spin–OAM information transfer and decoherence to high-energy interactions \cite{nagali2009:quantum,ndagano2017:characterizing}. 
Their controlled generation could therefore open new opportunities for nuclear-structure probes  \cite{klein2019:imaging,budker2022:expanding,Lu:2023wrf,balabanski2024:nuclear,thorsten2024:photoexcitation,borge2025:nuclear,liu2026:vortex}, spin-sensitive high-energy interactions \cite{aidala2013:spin,leader:2016photon,Ivanov:2019vxe,ivanov:2022promises}, and high-energy quantum information \cite{hiesmayr2024:quantum,bordes2024:first}.

At optical frequencies, vector-vortex beams can be generated directly via wavefront engineering, interferometric or quantum-interference schemes, and metasurfaces \cite{devlin:2017arbitrary,jimenez2017spontaneous,dorrah:2022tunable}. At shorter wavelengths, high-harmonic generation has produced extreme-ultraviolet vector-vortex beams \cite{delasHeras:22}, while free-electron lasers \cite{sasaki:2008proposal,bahrdt:2013first,rebernik2017:extreme} and laser--plasma interactions \cite{geng2026:generating} provide further routes to structured radiation. In the $\gamma$-ray regime, beam--target simulations predict cylindrical-vector polarization textures \cite{cao2025:generating,liu2026generation}. However, such polarization maps alone do not establish the intermode coherence required for single-photon spin--OAM nonseparability. 
Since conventional optics offer no straightforward route to shaping such $\gamma$-ray modes after emission, the central generation challenge is to establish and control their spin--OAM coherence directly at the radiation vertex.

Laser-induced nonlinear Compton scattering (NCS) has emerged as a promising source of bright, polarized $\gamma$ radiation \cite{li2020:polarized,mirzaie:2024all,wu:2025achieving}.
Angular-momentum transfer in Compton scattering and NCS enables vortex-photon generation \cite{jentschura2011:generation,petrillo:2016compton,chen:2018gamma,maruyama:2025photon}, with studies encompassing circularly and linearly polarized (LP) driving fields \cite{Ababekri:2022mob,Bogdanov2024:orbital,liao:2025all} and recent experimental evidence from an all-optical inverse-Compton source \cite{wei2026:experimental}.
Multifrequency and two-color driving fields further enable control of photon polarization and vortex charge \cite{Jiang:2024fit}, as well as coherent OAM superpositions within a fixed-helicity sector \cite{zhou2026:manipulation}, while electron wavepacket shaping provides an additional handle on the OAM spectrum \cite{zuo2026:photon}.
Vector-vortex emission, however, requires helicity--OAM channels such as $\ket{R,m}$ and $\ket{L,-m}$ to be populated coherently within the same selected photon--recoil kinematics, with controlled relative amplitudes and a well-defined relative phase.
The critical challenge is to identify and control multiphoton pathways with different angular-momentum transfer that coherently populate these opposite-helicity channels.

\begin{figure}[!t]
\includegraphics[width=1.0\linewidth]{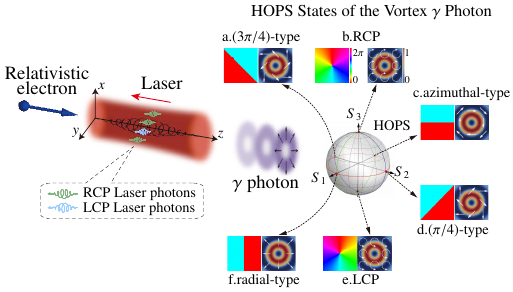}

\vspace{-10pt}
\caption{Generation of vector-vortex $\gamma$ photons via NCS.
A relativistic electron beam collides head-on with an arbitrarily polarized laser field, whose RCP and LCP components are shown in the inset.
Laser ellipticity controls their relative weights and hence the photon state on the higher-order Poincar\'e sphere (HOPS): circular polarization favors scalar-vortex states near the poles (b,e), whereas an LP drive balances the two helicity pathways and yields equatorial vector-vortex states, including radial-type (f), azimuthal-type (c), and hybrid modes (a,d).
Each HOPS panel shows transverse phase (left) and intensity (right), with arrows indicating the local polarization.}
\label{fig:1}
\end{figure}

\begin{figure*}
    \centering
    \includegraphics[width=1.0\linewidth]{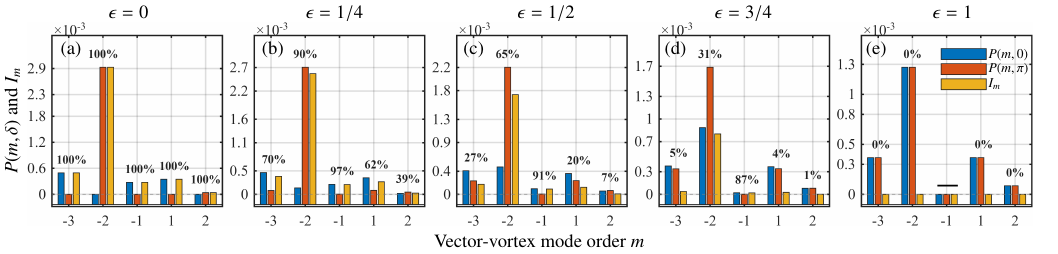}

    \vspace{-5pt}
    \caption{
Laser-ellipticity dependence of the integrated probabilities and interference magnitude for different vector-vortex mode orders $m$. 
The blue and orange bars show $P(m,0)$ and $P(m,\pi)$ for the
radial- and azimuthal-type analyzers, respectively, while the yellow
bars show the interference magnitude $I_m$. 
All quantities are integrated over
$0\leq\omega'\leq36~{\rm MeV}$ and
$0\leq\theta_{k'}\leq2.5~{\rm mrad}$. 
Panels (a)--(e) correspond to
$\epsilon=0,1/4,1/2,3/4,$ and $1$, respectively.
The percentages above the bars denote the concurrence $C_m$ of the corresponding mode pair.
}
    \label{fig:3}
\end{figure*}

\begin{figure}
    \centering
    \includegraphics[width=1.0\linewidth]{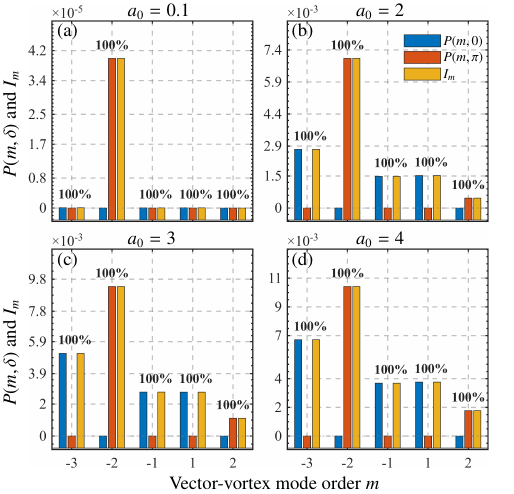}

    \vspace{-5pt}
    \caption{
Laser-intensity dependence of the integrated probabilities and interference magnitude for different vector-vortex mode orders $m$, for an LP laser field ($\epsilon=0$). The blue and orange bars show $P(m,0)$ and $P(m,\pi)$ for the radial- and azimuthal-type analyzers, respectively, while the yellow bars show the interference magnitude $I_m$. 
The same energy--angle integration window as in Fig.~\ref{fig:3} is used.
Panels (a)--(d) correspond to $a_0=0.1,2,3,$ and $4$, respectively.
The percentages above the bars denote the concurrence $C_m$ of the corresponding mode pair.
}
    \label{fig:4}
\end{figure}

In this Letter, we propose a novel method to generate vector-vortex $\gamma$ photons via NCS in elliptically polarized lasers (Fig.~\ref{fig:1}). 
Within strong-field quantum electrodynamics (QED) \cite{ritus1985quantum,fedotov2023:advances}, we show how the counter-rotating components of the driving field support multiphoton pathways coupling photon helicity to distinct OAM values, 
coherently populating the $\ket{R,m}$ and $\ket{L,-m}$ channels within the same selected photon--recoil kinematics. 
Tuning the laser ellipticity $\epsilon$ continuously rebalances these pathways, driving the emitted $\gamma$ photon between scalar- and vector-vortex configurations: circular polarization favors scalar vortex modes [Figs.~\ref{fig:1}(b) and (e)], whereas an LP field balances the two helicity--OAM arms, yielding radial- and azimuthal-type vector-vortex mode pairs [Figs.~\ref{fig:1}(c) and (f)] with pair concurrence approaching unity (Fig.~\ref{fig:3}). 
The normalized laser amplitude $a_0$ provides a secondary control knob by broadening the harmonic spectrum and accessible OAM-mode distribution (Fig.~\ref{fig:4}).
These controls provide a direct route to generating spin--OAM-entangled $\gamma$ photons, with potential applications in nuclear photonics and high-energy quantum information. Natural units ($\hbar=c=1$) are used throughout, with $m_e$ the electron mass and $\alpha=e^2/(4\pi)\approx1/137$.

\textit{Spin--OAM Encoding in Nonlinear Compton Emission.---}We treat the electron--laser interaction within the Furry picture using
Volkov states \cite{ritus1985quantum,fedotov2023:advances}. The laser potential is    
$A^\mu(\phi)=g(\phi)a_0m_e(0,\cos\phi,\epsilon\sin\phi,0)/
[e\sqrt{1+\epsilon^2}]$, where $\phi=k\cdot x$, $a_0$ is the dimensionless field amplitude, $g$ is the pulse envelope
and $\epsilon\in[0,1]$ varies the polarization from linear to circular.
Denoting the initial and final
electron momenta and helicities by $(p,\lambda)$ and $(p',\lambda')$, and
the emitted photon labels by $(k',\Lambda')$, the leading-order amplitude is
\begin{equation}
S_{fi}=-ie\int d^4x\,
\bar\psi_{p',\lambda'}(x)\slashed A_{k',\Lambda'}^{\prime *}(x)
\psi_{p,\lambda}(x).\tag{1}
\end{equation}
Details of its evaluation, including the
finite-pulse approximation to the linear Volkov phase, are given in
Appendix~\hyperref[app:A]{A} and Sec.~I of the Supplemental Material (SM) \cite{supplement_material}.

\begin{figure*}    
    \centering
    \includegraphics[width=1.0\linewidth]{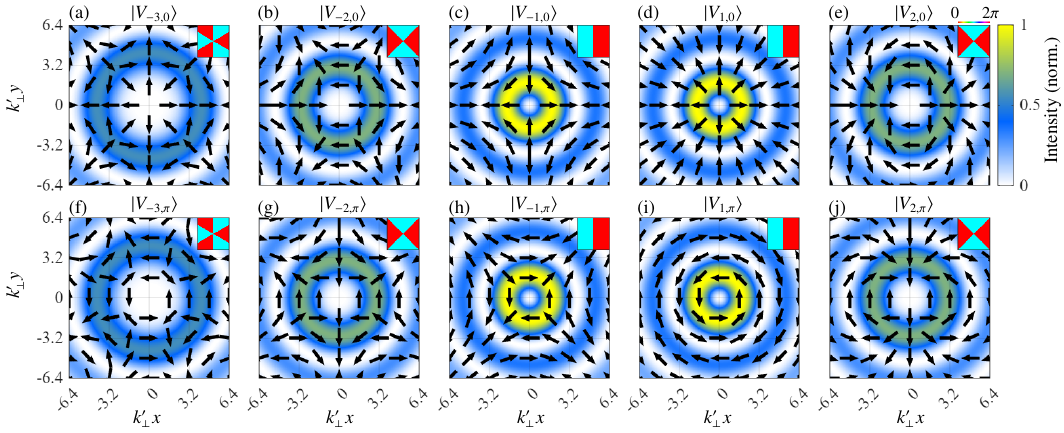}
    
    \vspace{-5pt}
    \caption{Transverse intensity and polarization distributions of representative vector-vortex \(\gamma\)-photon modes \(\ket{V_{m,\delta}}\) defined in Eq.~\eqref{eq:2}. Panels (a)-(e) show the radial-type modes (\(\delta=0\)), while panels (f)-(j) show the corresponding azimuthal-type modes (\(\delta=\pi\)), with mode orders \(m=-3,-2,-1,1\), and \(2\). The color maps represent the normalized transverse intensity distributions in the transverse coordinate plane, while the black arrows indicate the local linear-polarization directions.
    Insets show the corresponding transverse phase distributions. These normalized profiles illustrate the analyzer basis and do not represent calculated NCS emission distributions.}
    \label{fig:2}
\end{figure*}

NCS provides a route to photon states carrying
OAM through angular-momentum transfer in the electron--laser interaction
\cite{jentschura2011:generation,jentschura2011:compton,Ababekri:2022mob,Jiang:2024fit}. Such radiation can be described in a Bessel basis: each mode
is a coherent superposition of plane waves on a fixed momentum cone and
an eigenstate of axial total angular momentum (TAM). In the
relativistic cone considered here, $\theta_{k'}\sim1/\gamma_0\ll1$ for
$a_0$ of order unity, with $\gamma_0=\varepsilon/m_e$, the paraxial mode
is described to leading order by a scalar OAM factor and a helicity
polarization vector. We label these components $\ket{\Lambda',\ell'}$,
where $\ell'\in\mathbb Z$, $\Lambda'=\pm1$, and $j=\ell'+\Lambda'$.
This leading-order separation provides a basis for describing correlations
between the emitted photon's helicity and OAM.

An elliptically polarized laser contains two counter-rotating circular
components with relative amplitudes proportional to $1+\epsilon$ and
$1-\epsilon$. Their oppositely signed spin-angular-momentum transfer provides distinct multiphoton pathways that can populate opposite OAM sectors of the emitted $\gamma$ photon.
For a fixed electron-spin channel and recoil-mode projection, the
conditional photon amplitude can be expanded as
$\ket{\Psi_\gamma^{\lambda'\lambda}}\propto
\sum_{\Lambda',\ell'}c_{\Lambda',\ell'}^{\lambda'\lambda}
\ket{\Lambda',\ell'}$, with the continuous photon labels suppressed (Appendix~\hyperref[app:B]{B}).
When opposite emitted helicities coherently populate opposite OAM modes,
a selected pair can take the two-mode vector-vortex form $c_{R,m}|R,m\rangle + c_{L,-m}|L,-m\rangle$ with
nonseparable helicity and OAM for $m\ne0$ and two nonzero amplitudes.
An equal-weight pure pair lies on the higher-order Poincar\'e-sphere (HOPS)
equator and has the  form
\begin{equation}
\ket{V_{m,\delta}}=\frac{\ket{R,m}+e^{i\delta}\ket{L,-m}}{\sqrt2}. 
\tag{2}\label{eq:2}
\end{equation}
The order $m$ and relative phase $\delta$ label this analyzer family; the symmetric equal-weight family in Eq.~\eqref{eq:2} is the opposite-OAM subset of generalized two-mode vector-vortex states defined in Appendix~\hyperref[app:B]{B}. 
Under this convention, $\delta=0$ yields the radial-type state [Fig.~\ref{fig:1}(f)] and $\delta=\pi$ yields the azimuthal-type state [Fig.~\ref{fig:1}(c)], while intermediate phases generate hybrid textures [Figs.~\ref{fig:1}(a) and (d)]. 
The photon-mode phases are fixed by the common convention in Sec.~III of the SM \cite{supplement_material}.
Laser ellipticity controls the relative pathway weights, while their
coherence determines the vector-vortex character of the selected pair.
Unresolved electron spins and photon kinematics are combined at the
density-matrix level, as specified in Appendix~\hyperref[app:C]{C}.

To define this conditional state, we project the recoil electron onto an
axially symmetric zero-vortex mode \cite{karlovets2022:generation}. This is a coherent mode projection;
an incoherent acceptance of recoil momenta need not preserve the same
two-arm coherence. Finite Gaussian recoil modes and Bessel--Gauss photon
packets provide normalizable transverse descriptions (Sec.~II of the SM \cite{supplement_material}).
Their common-angle integration retains the sector-wise angular matching,
while the radial and relative-angle overlaps determine the mode weights
and visibility.

With the differential prefactor included in coefficients $c$, the energy--angle resolved probability density for $\gamma$ photon emission into the vector-vortex analyzer channel for an unpolarized incident electron is 
\begin{equation}
\frac{d^2W^{\rm VV}_{m,\delta}}{d\omega'\,d\theta_{k'}}
=\frac12\sum_{\lambda,\lambda'}
\left|\frac{c_{R,m}^{\lambda'\lambda}
       +e^{-i\delta}c_{L,-m}^{\lambda'\lambda}}{\sqrt2}\right|^2
=B_m+\operatorname{Re}(e^{-i\delta}\Gamma_m),\tag{3} \label{eq:3}
\end{equation}
where the final-spin sum is over the accepted channels,
$B_m=(P_{R,m}+P_{L,-m})/2$, and
$\Gamma_m=\frac12\sum_{\lambda,\lambda'}
(c_{R,m}^{\lambda'\lambda})^*c_{L,-m}^{\lambda'\lambda}$;
the populations use the same spin average. The calculations below retain the non-spin-flip channel, $\lambda'=\lambda$, which dominates for the moderate-field regime $a_0\sim1$ considered here. The interference term makes the relative two-arm
coherence observable through the analyzer phase. The comparison with
classical-current calculations in Sec.~IV of the SM \cite{supplement_material} reproduces the
principal normalized spectral structures and analyzer enhancement and
suppression, while retaining a visible displacement of the higher-energy band. 
Thus, the two twisted-photon components and their analyzer interference can already arise from coherent radiation along an electron trajectory.

For numerical demonstrations, we consider a head-on collision between an ultrarelativistic  electron beam (\(\varepsilon=1~{\rm GeV}\)) and a finite laser pulse of central 
photon energy \(\omega=1.55~{\rm eV}\) ($\lambda_0 \simeq 0.8~\mu\text{m}$) and a $\cos^2$ envelope ($N_{\text{cycle}} = 10$, full-support duration $\tau \simeq 26.7$~fs). Figure~\ref{fig:3} uses the baseline normalized amplitude $a_0 = 1$, corresponding to a peak intensity of order $10^{18}~\text{W/cm}^2$. The intensity scaling and full $a_0$ dependence are evaluated in Fig.~\ref{fig:4}.

{\textit{Vector-vortex analyzer modes.---}}Figure~\ref{fig:2} illustrates the real-space intensity and polarization textures of representative vector-vortex analyzer modes defined in Eq.~\eqref{eq:2}, including radial-type [\(\delta=0\), see Figs.~\ref{fig:2}(a)-(e)] and azimuthal-type [\(\delta=\pi\), see Figs.~\ref{fig:2}(f)-(j)] states.
These profiles characterize the chosen mode basis; their NCS excitation probabilities and coherence are calculated below. 
Each mode exhibits the hallmark central intensity null of nonzero winding numbers alongside a spatially varying polarization.  
The order $m$ labels the pair $\{\ket{R,m},\ket{L,-m}\}$. For $m=-1$ [Figs.~\ref{fig:2}(c),(h)], both arms have zero axial TAM; the other displayed pairs combine the sectors $j=\pm(m+1)$, directly encoding an intrinsic spin--OAM nonseparability at the emission vertex.

The NCS excitation of these analyzer modes is governed by angular-momentum transfer.
The  indices $n_1$ and $n_2$ label the two Fourier/Bessel channel families,  with $q=n_1-n_2$ their net laser-helicity index. TAM conservation enforces $\ell'+h_{\Lambda'} = q + \lambda - \lambda'$, which reduces to $\ell' = q - h_{\Lambda'}$ for the retained spin-conserving channels ($\lambda = \lambda'$). The emitted photon OAM is thus uniquely fixed by the net laser helicity minus the photon's own SAM.

A vector-vortex pair arises when the two photon-helicity amplitudes coherently populate the \(\ket{R,m}\) and \(\ket{L,-m}\) arms. In the retained spin-conserving channels, their net-helicity indices are \(q_R = m+1\) and \(q_L = -(m+1)\). These are opposite sectors for \(m\neq -1\); for \(m = -1\), both arms instead belong to \(q = 0\). An LP drive (\(\epsilon = 0\)) contains equal circular components, and the calculated selected pairs have balanced, coherent amplitudes. For the recoil-mode projection specified above, 
these pathways interfere coherently upon projection onto the common analyzer, yielding the detected amplitude $\mathcal A^{\rm VV}_{m,\delta} = (c_{R,m} + e^{-i\delta}c_{L,-m})/\sqrt{2}$ without $\gamma$-ray wavefront shaping after
emission. Radial-type ($\delta=0$) and azimuthal-type ($\delta=\pi$) modes thus select different analyzer phases [Figs.~\ref{fig:1}(c),(f) and~\ref{fig:3}(a)].
In the CP limit \(\epsilon = 1\), one circular component of the drive is absent, and the resolved selected opposite-OAM pairs have vanishing interference, as quantified in Fig.~\ref{fig:3}(e).

\textit{Laser-ellipticity-controlled spin--OAM nonseparability.---}To quantify how laser ellipticity $\epsilon$ controls the population and coherence of different vector-vortex mode pairs, we integrate the energy--angle resolved probability density in Eq.~\eqref{eq:3}, obtaining the analyzer-resolved integrated probability $P(m,\delta)$ for each mode order $m$. 
Over the same window, the interference magnitude
$I_m=2|\Gamma_m|$ quantifies the coherence between
$\ket{R,m}$ and $\ket{L,-m}$.
Since
$C_m=I_m/(P_{R,m}+P_{L,-m})$, stronger relative interference corresponds to greater spin--OAM nonseparability within the selected mode pair, with $C_m\to1$ for an ideal vector-vortex pair. 
Figure~\ref{fig:3} shows that varying $\epsilon$ redistributes both the population and coherence among the different mode pairs.
Importantly, $C_m$ characterizes the internal spin--OAM nonseparability of a selected pair, not its weight in the full multimode emission. A large \(C_m\) therefore does not imply that the selected pair dominates the full conditional radiation.

For an LP drive ($\epsilon=0$), the two laser helicity resources are perfectly balanced. The interference magnitude \(I_m\) reaches the maximum allowed by the corresponding pair populations, yielding $C_m \simeq 1.00$ for the orders in Fig.~\ref{fig:3}(a). This confirms that, within each selected HOPS pair, the two arms are coherently balanced. The emission strength of each order is determined separately by the corresponding probability bars.

As $\epsilon$ increases, the changes in population balance and coherence are strongly mode dependent. The $m=-1$ pair retains a large displayed concurrence at intermediate ellipticities, even when it is not the most strongly populated pair. At $\epsilon=1/4$ [Fig.~\ref{fig:3}(b)], the negative orders $m=-3,-2,-1$ retain sizable concurrence ($C_m \simeq 70\%, 90\%, 97\%$), whereas the positive-$m$ modes already show a substantial reduction. At intermediate ellipticity \(\epsilon=1/2\) [Fig.~\ref{fig:3}(c)], the concurrence becomes strongly mode dependent, with the \(m=-1\) channel retaining a concurrence of 91\% while other modes are progressively suppressed. By $\epsilon=3/4$ [Fig.~\ref{fig:3}(d)], the drive imbalance suppresses most modes strongly, leaving the \(m=-2\) channel as the dominant contribution with two-arm coherence. In the circular limit $\epsilon=1$ [Fig.~\ref{fig:3}(e)], one driving-field helicity component is absent, and the resolved opposite-OAM pairs have vanishing interference and concurrence. The dash for \(m=-1\) indicates an unresolved pair population, for which a reliable conditional concurrence is unavailable.

The sequence in Fig.~\ref{fig:3} demonstrates a continuous transition from balanced spin--OAM pairs at $\epsilon=0$ to vanishing concurrence in the resolved selected pairs at $\epsilon=1$. Ellipticity therefore acts as the primary topological control parameter, directly tuning the relative magnitudes of $c_{R,m}$ and $c_{L,-m}$ to shift the HOPS position of the selected pair.  Meanwhile, the TAM rule selects the net laser helicity index $q=n_1-n_2$, while the pulse profile organizes the allowed channels across energy and angle. The energy--angle spectra in Appendix~\hyperref[app:D]{D} provide complementary kinematic fingerprints of these multiphoton channel families to help identify favorable mode orders, though full verification of vector-vortex nonseparability fundamentally relies on resolving the helicity--OAM coherence.

\textit{Intensity scaling and experimental outlook.---}While ellipticity controls the relative circular components, $a_0$
controls the spectral and harmonic content. Figure~\ref{fig:4} shows
$P(m,0)$, $P(m,\pi)$, and $I_m$ over the same energy--angle window as
Fig.~\ref{fig:3} for an LP drive ($\epsilon=0$) at
$a_0=0.1,2,3,4$. As $a_0$ increases, additional multiphoton channels
are activated, broadening the accessible mode spectrum and enhancing
the emission strength. 
Meanwhile, $I_m$ remains close to $P_{R,m}+P_{L,-m}$ for the
resolved pairs, keeping $C_m$ near unity and preserving their
vector-vortex character.

The calculations demonstrate control of the selected spin–OAM coherence across the laser-parameter range considered here. The required electron-energy scale is experimentally accessible: hundred-MeV electrons have been used in inverse-Compton vortex $\gamma$-ray studies \cite{wei2026:experimental}, and laser-wakefield acceleration to 1 GeV has been demonstrated \cite{leemans:2006gev}. Together with all-optical nonlinear Compton experiments \cite{mirzaie:2024all}, these results motivate testing the proposed emission geometry. Verification of the vector-vortex character requires sensitivity to helicity--OAM coherence under the recoil conditioning assumed here. The predicted analyzer contrasts provide targets for future polarization- and mode-sensitive measurements. 
Although these measurements remain experimentally challenging, they provide a concrete route for exploiting this dual control—$\epsilon$ for topology and $a_0$ for bandwidth—to test spin--OAM structured radiation in upcoming strong-field QED campaigns. 

In conclusion, we have shown that NCS provides a route to conditional vector-vortex $\gamma$ photons, bridging structured optics and high-energy radiation. By tuning the laser ellipticity $\epsilon$, we demonstrate a continuous transition from scalar-vortex states to maximally nonseparable selected vector-vortex mode pairs, with the degree of spin--OAM entanglement controlled by coherent interference between competing multiphoton pathways. The laser amplitude further controls the accessible OAM-mode spectrum. Crucially, this encoding occurs directly at the radiation vertex, without requiring post-emission $\gamma$-ray phase optics. These results extend structured photonics into the strong-field QED regime at MeV energies and open prospects for spin--OAM-sensitive high-energy interactions and high-energy quantum information. \\

{\it Acknowledgments---}This work is supported by the National Natural Science Foundation of China (Grants No. 12425510, No. U2267204, No. 12441506, No. 12505276, No. 12447106, No. 12475249), the National Key Research and Development (R\&D) Program (Grant No. 2024YFA1610900), the Science Challenge Project (No. TZ2025012), the Innovative Scientific Program of CNNC, and the Fundamental Research Funds for the Central Universities (No. xzy012026064).

\bibliographystyle{apsrev4-2-year}
\bibliography{apssamp}

\onecolumngrid

\vspace{2.0\baselineskip}

\twocolumngrid

\section*{Appendix A: The $S$-Matrix Element} \label{app:A}


The leading-order $S$-matrix element for NCS in an arbitrarily polarized laser pulse can be written as
\begin{equation}
S_{fi}
= ie\,\frac{(2\pi)^3}{\omega}
\delta^3(\bm p+s_0\bm k-\bm p'-\bm k')
\sum_{\bm n}
e^{i \mu_{\bm n}\varphi_{p'}}
\mathcal M_{\bm n}^{\Lambda'}(s_0),
\label{eq:Sfi_app} \tag{A1}
\end{equation}
where $\bm n\equiv(n_1,n_2)$ and
$\mu_{\bm n}=n_1-n_2+\lambda-\lambda'$. The multiphoton amplitude expands in Wigner $d$-functions as
\begin{equation}
\begin{aligned}
\mathcal M_{\bm n}^{\Lambda'}(s)={}&
d^{1/2}_{\lambda,\lambda'}d^1_{0,\Lambda'}
\widetilde{\mathcal G}_0^{\uparrow\uparrow}
+e^{-i2\lambda\Delta\varphi}d^{1/2}_{-\lambda,\lambda'}
d^1_{2\lambda,\Lambda'}\widetilde{\mathcal G}_0^{\uparrow\downarrow}\\
&+d^{1/2}_{\lambda,\lambda'}
\left[e^{-i2\lambda\Delta\varphi}d^1_{2\lambda,\Lambda'}
\widetilde{\mathcal G}_{\pm1,1}^{\uparrow\uparrow}
+e^{i2\lambda\Delta\varphi}d^1_{-2\lambda,\Lambda'}
\widetilde{\mathcal G}_{\pm1,2}^{\uparrow\uparrow}\right]\\
&+d^{1/2}_{-\lambda,\lambda'}d^1_{0,\Lambda'}
\widetilde{\mathcal G}_{\pm}^{\uparrow\downarrow},\\
\end{aligned}
\label{eq:M_app} \tag{A2}
\end{equation}
with $\Delta\varphi=\varphi_{p'}-\varphi_{k'}$.  
The auxiliary functions $\widetilde{\mathcal G}$ are defined in Sec. I of SM \cite{supplement_material}. Equations~\eqref{eq:M_app} and ~\eqref{eq:Sfi_app} serve as the input for deriving the emitted-photon OAM selection rules.


\section*{Appendix B: Bessel projection and vector-vortex analyzer}\label{app:B}

We project the emitted radiation onto a Bessel photon mode about the collision axis, which is an eigenstate of axial TAM. In the ultrarelativistic NCS geometry considered here (\(\theta_{k'}\sim1/\gamma_0\ll1\)), it reduces at leading paraxial order to
\(|\Lambda',\ell'\rangle\leftrightarrow\chi_{\Lambda'}e^{ik_z'z}e^{i\ell'\varphi}J_{\ell'}(k'_\perp r_\perp)\).
With \(h_R=+1\) and \(h_L=-1\), its axial angular momentum is \(J_z^{(\gamma)}=\ell'+h_{\Lambda'}\).

For the recoil electron we use a cylindrically symmetric zero-vortex mode,
\(
|\ell_e=0;p'_\perp,p'_z,\lambda'\rangle
\propto
\int_0^{2\pi}d\varphi_{p'}
|\bm{p}'(\varphi_{p'}),\lambda'\rangle 
\), which fixes the transverse momentum modulus but carries no
azimuthal phase. It serves as the minimal paraxial, axially symmetric recoil-mode resolution used to expose angular-momentum transfer, rather than as an additional topological filter. A finite
cylindrically symmetric Gaussian recoil acceptance gives the same
azimuthal selection rule, with the radial delta function replaced by a
finite overlap integral (Sec. II of SM \cite{supplement_material}).

Combining the azimuthal phase in Appendix~\hyperref[app:A]{A} with the Bessel photon and
cylindrical recoil projections yields
$\int_0^{2\pi} d\varphi\,e^{i(\mu_{\bm n}-\ell'-h_{\Lambda'})\varphi}=2\pi\,\delta_{\ell'+h_{\Lambda'},\,\mu_{\bm n}}$, hence 
$\ell' = n_1-n_2+\lambda-\lambda'-h_{\Lambda'}$.
In the spin-conserving regime ($\lambda=\lambda'$), this reduces to $\ell' = n_1-n_2-h_{\Lambda'}$, meaning the emitted photon OAM is the net laser helicity absorbed by the channel minus the emitted photon helicity.

After the Bessel/TAM projection, the emitted photon state at fixed
\((\omega',\theta_{k'})\) expands as
\[
|\Psi_\gamma\rangle
=
\sum_{\Lambda'=R,L}
\sum_{\ell'\in\mathbb Z}
c_{\Lambda',\ell'}^{\lambda'\lambda}
|\Lambda',\ell'\rangle ,
\]
where the coefficients are given by
\begin{equation}
c_{\Lambda',\ell'}^{\lambda'\lambda}
=\sqrt{\frac{\alpha k'_\perp}{(k\cdot p)(k\cdot p')}}\,
h_{\Lambda'}(-i)^{\ell'+h_{\Lambda'}}
\sum_{n_1,n_2\in\mathbb Z}
\delta_{\ell'+h_{\Lambda'},\mu_{\bm n}}
\mathcal M_{\bm n}^{\Lambda'}. \tag{B1}
\end{equation}

The vector-vortex analyzer of order \(m\) is defined as
\begin{equation}
  |V_{m,\delta}\rangle
=
\frac{1}{\sqrt2}
\left(
|R,m\rangle+e^{i\delta}|L,-m\rangle
\right),\label{vv_balanced}  \tag{B2}
\end{equation}
where the relative phase \(\delta\) determines the orientation of the
spatial polarization texture ($\delta=0$ yields a radial-type state and $\delta=\pi$ yields an azimuthal-type state). 
The structural overlap with a scalar spin--OAM basis state is
\(
\mathcal F_{\rm struct}
\equiv
\langle V_{m,\delta}|\Lambda',\ell'\rangle
=
\frac{1}{\sqrt2}
\left(
\delta_{\Lambda',R}\delta_{\ell',m}
+
e^{-i\delta}\delta_{\Lambda',L}\delta_{\ell',-m}
\right)
\).

The detected vector-vortex amplitude can be written as a projection of the evolved photon state,
\[
\mathcal A^{\rm VV}_{m,\delta}
=
\sum_{\Lambda',\ell'}
\mathcal F_{\rm struct}\,
c_{\Lambda',\ell'}^{\lambda'\lambda}
=
\frac{1}{\sqrt2}
\left(
c_{R,m}^{\lambda'\lambda}+e^{-i\delta}c_{L,-m}^{\lambda'\lambda}
\right).
\]
Thus, the two analyzer arms select \(\mu_{\bm n}=m+1\) for \(|R,m\rangle\) and \(\mu_{\bm n}=-m-1\) for \(|L,-m\rangle\).

The opposite-OAM family used in Eq.~\eqref{vv_balanced} is a minimal symmetric choice. More
generally, a normalized pure two-mode state is \cite{milione2011:higher,Forbes:2021tpp}
\begin{equation}
 |\psi\rangle=\cos\frac{\vartheta}{2}|R,\ell_a\rangle
   +e^{i\delta}\sin\frac{\vartheta}{2}|L,\ell_b\rangle,
 \qquad 0\leq\vartheta\leq\pi. \tag{B3}
\end{equation}
Here $\vartheta$ is the HOPS polar angle, not the emission angle.
For $\ell_a\ne\ell_b$ and two nonzero coefficients, helicity and OAM are
nonseparable. We choose $\ell_a=m$, $\ell_b=-m$, with $m\ne0$, so that
the paraxial Bessel components have equal radial intensity profiles at
the same cone momentum: $|J_m(k'_\perp r_\perp)|^2
=|J_{-m}(k'_\perp r_\perp)|^2$. Unequal absolute OAM indices generally
give different radial profiles but do not invalidate the two-mode state.

\begin{figure*}
    \centering
    \includegraphics[width=1.0\linewidth]{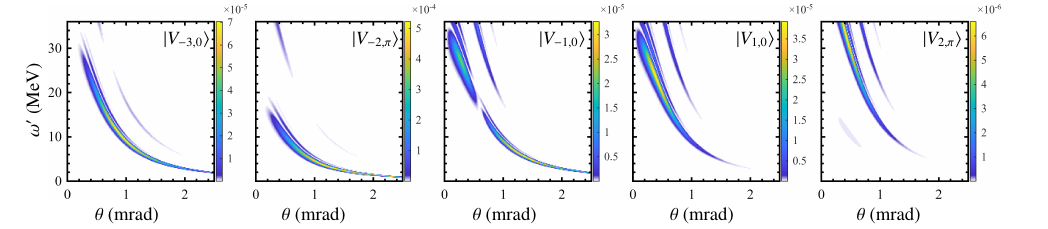}
    \vspace{-20pt}
    \caption{Energy–angle probability-density distributions for selected radial-type (\(m=-3,-1,1\)) and azimuthal-type (\(m=-2,2\)) vector-vortex \(\gamma\)-photon modes produced via NCS in an LP laser field. Note that the color-bar scale differs among panels.}
    \label{fig:D1}
\end{figure*}

\section*{Appendix C: Concurrence, pair weight, and HOPS coordinates}  \label{app:C}

After conditioning on the selected recoil mode and tracing over unresolved
electron spins and photon kinematics in the acceptance $\mathcal A$, the
normalized density matrix of the selected pair is
\begin{equation}
\rho_m=\frac1{W_m}
\begin{pmatrix}P_{R,m}&\Gamma_m^*\\\Gamma_m&P_{L,-m}\end{pmatrix},
\qquad W_m=P_{R,m}+P_{L,-m}. \tag{C1}
\end{equation}

Writing $a$ for the retained electron-spin channels and $w_a$ for their
statistical weights, its unnormalized entries are
\begin{equation}
\begin{aligned}
P_{R,m}&=\sum_a w_a\int_{\mathcal A}d\omega'\,d\theta_{k'}
 |c_{R,m;a}|^2,\\
P_{L,-m}&=\sum_a w_a\int_{\mathcal A}d\omega'\,d\theta_{k'}
 |c_{L,-m;a}|^2,\\
\Gamma_m&=\sum_a w_a\int_{\mathcal A}d\omega'\,d\theta_{k'}
 c_{R,m;a}^*c_{L,-m;a}. 
\end{aligned}\tag{C2}
\end{equation}
For the present spin-conserving calculation, $a=\lambda=\pm1/2$ and
$w_a=1/2$. The same expressions without the kinematic integrals define a
fixed-kinematics pair. For $m\ne0$, projection onto
$\{\ket{R,m},\ket{L,-m}\}$ gives the spin--OAM concurrence \cite{wootters1998:entanglement}
\begin{equation}
C_m=\frac{2|\Gamma_m|}{W_m}
\leq\frac{2\sqrt{P_{R,m}P_{L,-m}}}{W_m}\leq1. \tag{C3}
\end{equation}
A pure, fully coherent pair saturates the first bound. A vanishing or
numerically unresolved pair weight does not support a quantitative
conditional concurrence. The off-diagonal element is integrated before
taking its magnitude. Concurrence quantifies nonseparability, not
density-matrix purity.

To distinguish this pair nonseparability from its overall dominance in the full multimode emission, we define the pair weight as
\(\eta_m^{\rm pair}=(P_{R,m}+P_{L,-m})/\sum_{\Lambda',\ell'}P_{\Lambda',\ell'}\).
For the full conditional photon state to approximate one maximally nonseparable mode pair, both \(C_m\simeq1\) and \(\eta_m^{\rm pair}\simeq1\) are required. 

The same pair can be represented on HOPS via
the normalized Stokes-like coordinates
\begin{align}
	S_1^{(m)} &= \frac{2{\rm Re}\,\Gamma_m}{P_{R,m}+P_{L,-m}}, \nonumber\\
	S_2^{(m)} &= \frac{2{\rm Im}\,\Gamma_m}{P_{R,m}+P_{L,-m}}, \nonumber\\ \tag{C4}
	S_3^{(m)} &= \frac{P_{R,m}-P_{L,-m}}{P_{R,m}+P_{L,-m}}.
    \label{HOPS_coordinate}  
\end{align}
These coordinates obey $|\bm S^{(m)}|\leq1$: pure states lie on the HOPS
surface, while mixtures lie inside it. The equal-weight pure states
$|V_{m,\delta}\rangle$ lie on the equator, unequal coherent arm weights
give off-equator pure states, and the poles correspond to scalar
helicity--OAM product modes. The driving laser ellipticity controls the
final HOPS position by dynamically tuning the balance between the
underlying circular absorption pathways.

Equal populations alone imply $S_3^{(m)}=0$, but not maximal
nonseparability, which additionally requires $2|\Gamma_m|=W_m$.
For any fixed reference phase $\delta_0$, the orthogonal pair
$\{|V_{m,\delta_0}\rangle,|V_{m,\delta_0+\pi}\rangle\}$ forms a complete
basis of this two-mode subspace. Pure states are expanded coherently in
this basis, whereas mixed states require a density-matrix description.

\section*{Appendix D: Energy-Angle Kinematic Fingerprints} \label{app:D}

Each vector-vortex analyzer samples a distinct subset of multiphoton channels (Fig.~\ref{fig:D1}). While $s_0$ sets the energy--angle kinematics, the TAM condition ($\ell'+h_{\Lambda'}=n_1-n_2+\lambda-\lambda'$) selects the allowed net-helicity channels, producing distinct patterns in the $(\omega',\theta_{k'})$ plane across different $m$ orders.

These spectra serve as kinematic maps for selecting optimal detection windows. However, full verification of spin--OAM nonseparability requires resolving the off-diagonal helicity--OAM phase coherence via polarization-sensitive or recoil-resolved measurements.

\end{document}